# Fe-vacancy order and superconductivity in tetragonal $\beta$-Fe$_{1-x}$Se

Ta-Kun Chen[1*], Chung-Chieh Chang[1], Hsian-Hong Chang[2], Ai-Hua Fang[1], Chih-Han Wang[1], Wei-Hsiang Chao[1], Chuan-Ming Tseng[1], Yung-Chi Lee[1,3], Yu-Ruei Wu[1], Min-Hsueh Wen[1], Hsin-Yu Tang[4,5], Fu-Rong Chen[1,4], Ming-Jye Wang[1,2], Maw-Kuen Wu[1,3,6*], Dirk Van Dyck[5]

1. Institute of Physics, Academia Sinica, Taipei 115, Taiwan

2. Institute of Astronomy and Astrophysics, Academia Sinica, Taipei 115, Taiwan

3. Department of Physics, National Tsing Hua University, Hsinchu 300, Taiwan

4. Department of Engineering and System Science, National Tsing Hua University, Hsinchu 300, Taiwan

5. EMAT, Department of Physics, University of Antwerp, 2020 Antwerp, Belgium

6. Department of Physics, National Dong Hwa University, Hualien 974, Taiwan

To whom correspondence may be addressed. Ta-Kun Chen and Maw-Kuen Wu, e-mail: tkchen@phys.sinica.edu.tw and mkwu@mail.ndhu.edu.tw

**Abstract**

Several superconducting transition temperatures in the range of 30–46 K were reported in the recently discovered intercalated FeSe system ($A_{1-x}Fe_{2-y}Se_2$, A = K, Rb, Cs, Tl). Although the superconducting phases were not yet conclusively decided, more than one magnetic phase with particular orders of iron vacancy and/or potassium vacancy were identified, and some were argued to be the parent phase. Here we show the discovery of the presence and ordering of iron vacancy in nonintercalated FeSe (PbO-type tetragonal $\beta$-$Fe_{1-x}Se$). Three types of iron-vacancy order were found through analytical electron microscopy, and one was identified to be nonsuperconducting and magnetic at low temperature. This discovery suggests that the rich-phases found in $A_{1-x}Fe_{2-y}Se_2$ are not exclusive in Fe-Se and related superconductors. In addition, the magnetic $\beta$-$Fe_{1-x}Se$ phases with particular iron-vacancy orders are more likely to be the parent phase of FeSe superconducting system, instead of the previously assigned $\beta$-$Fe_{1+\delta}Te$.

**Significance Statement**

We report the observation in the FeSe system of various Fe-vacancy orders, which are similar to those discovered in $A_{1-x}Fe_{2-y}Se_2$ superconductors. We observed at least three different Fe-vacancy orders, and one was identified to be nonsuperconducting and magnetic at low temperature, which implies a rich-phases nature of Fe-based superconductors. Consequently, we propose the existence of a modified phase diagram for the Fe-Se superconducting system. This discovery provides new opportunities to investigate the correlation between superconductivity and the Fe-vacancy order, which is critical to understand the origin of superconductivity of FeSe and related superconductors.

\body

The iron pnictide superconductors have opened the door to a new way to obtain superconductivity at very high temperatures. $\beta$-$Fe_{1+\delta}Se$ is remarkable among those superconductors in that it contains the essential electronic and structural constituents required for superconductivity without the conceptual complexity seen in other systems (1). Previous studies showed that the superconducting property of β-$Fe_{1+\delta}$Se is very sensitive to its stoichiometry (1, 2). In Fe-Se binary phase diagram (2-4), the PbO-type tetragonal structure (the *β* phase) only stabilized at Fe-rich side ($\delta$ = 0.01–0.04), whereas bulk superconductivity was observed in samples with δ close to 0.01 (5). McQueen *et al.* showed no superconductivity for samples with δ = 0.03 (5). On the other hand, the fact that only one superconducting phase has been reported in FeSe, unlike the other Fe-As–based superconductors that exhibit clear doping dependence of superconductivity and the absence of superconductivity in FeTe, led to the suggestion that FeTe is the nonsuperconducting parent compound of FeSe (6). Thus, the phase diagram derived from this picture shows very different features compared with other Fe-As–based superconductors (6, 7). In this work we use low-temperature synthesis methods to prepare *β*-$Fe_{1-x}$Se for a wide range of compositions, which allows for the determination for the composition-dependent electronic behavior for this important superconducting system.

The recent discovered alkali/alkaline-intercalated iron selenide ($A_{1-x}Fe_{2-y}Se_2$) superconductors with rich superconducting phases, where A = K, Rb, Cs, Tl, attracted great attention not only due to its high superconducting transition temperature ($T_c$, up to 46 K) (8), but also because of their dissimilar characteristics as compared to other iron-based superconductors, especially its seemingly intrinsic multiphase nature and the presence of iron vacancies and orders in the nonsuperconducting regime (9-13). The most frequently observed Fe-vacancy order in $A_{1-x}Fe_{2-y}Se_2$ is the $\sqrt{5} \times \sqrt{5} \times 1$ superstructure, which yields a phase of $A_{0.8}Fe_{1.6}Se_2$ (or $A_2Fe_4Se_5$). Scanning tunneling microscopy (STM) (11, 14, 15) and transport studies (12, 13, 16, 17) showed that $A_2Fe_4Se_5$ is an antiferromagnetic (AFM) insulator. Neutron scattering measurements (9) revealed a blocked checkerboard AFM with magnetic moments along the *c* axis for $A_2Fe_4Se_5$, ordered at a temperature as high as >500 K, with an unexpected large ordered magnetic moment of ~3.3 $\mu_B$/Fe at 10 K. Experiments have further shown that the type of vacancy and magnetic orders is highly sensitive to the stoichiometry ($x$ and $y$) of $A_{1-x}Fe_{2-y}Se_2$. Reports have shown the existence of other Fe-vacancy order with the forms $\sqrt{2} \times \sqrt{2} \times 1$ (10), $\sqrt{2} \times 2\sqrt{2} \times 1$ (13, 18), and $\sqrt{8} \times \sqrt{10} \times 1$ (19). However, the magnetic properties such as the type and transition temperature of the magnetic order are far less studied compared with that of the $K_2Fe_4Se_5$ phase. In addition, there were also results showing in $K_{1-x}Fe_{2-y}Se_2$ samples with a typical $T_c = 31$ K and additional superconducting phase with $T_c = 44$ K (20), whereas no clear identification of the new phases was available.

The complexity of phases and phase separation during crystal preparation in $A_{1-x}Fe_{2-y}Se_2$ make it difficult to conclusively verify the phase-property relationship, even for the superconducting phases. $\beta$-$Fe_{1+\delta}Se$, on the other hand, has the simplest structure among all iron-based superconductor families. Several surprising results related to the Fe-Se system appeared in the literature during the last few years, including the enhancement of $T_c$ to about 40 K under high pressure (21-23) and the intriguing extremely high $T_c$ (with a superconducting energy gap of ~20 meV) in molecular beam epitaxy (MBE)-grown single-layer FeSe (24-26). We also demonstrated the presence of superconducting-like feature with $T_c$ close to 40 K in samples of nano-dimensional form (27). Therefore, it is quite natural to ask whether the presence of the complex phases observed in $A_{1-x}Fe_{2-y}Se_2$ compounds and Fe-vacancy order exist in samples without alkaline metals. Here we present the first discovery of iron vacancies and three types of vacancy orders in tetragonal $\beta$-$Fe_{1-x}Se$, characterized by analytical transmission electron microscopy (TEM). Our observations imply that a unprecedented phase diagram should be considered in the Fe-Se superconductors.

**Results and Discussion**

Figure 1(a) is the TEM image of a FeSe nanowire, which exhibits the tetragonal symmetry identified from selected-area electron diffraction (SAED) shown in Fig. 1(b). In contrast to the expectation, electrical resistivity measurements on these nanowires showed that they were not superconducting, as displayed in the inset of Fig. 1(a). To understand the reason for not observing superconductivity, we performed more detailed electron diffraction of the nanowires. Figure 1(b) shows the SAED pattern taken from a tetragonal FeSe nanowire, with clear superstructure spots lay on the reciprocal $a^*$-$b^*$ plane that could be unambiguously identified as the $\sqrt{5} \times \sqrt{5} \times 1$ Fe-vacancy order by the unique superstructure wave vector $\mathbf{q}_1 = (1/5, 3/5, 0)$. This order is exactly the same as the one found in $A_2Fe_4Se_5$. We then denominated this phase as $\beta$-$Fe_4Se_5$ ($x = 0.2$). This observation resolved our puzzle of the absence of superconductivity in these FeSe nanowires. A subsequent question was whether this Fe-vacancy order exists in other FeSe materials? We decided to examine the other FeSe samples we have synthesized, including particles and sheets prepared using hydrothermal process (27), crystals grown at high pressure (28), and samples from $K_2Fe_4Se_5$ crystals after extracting K by iodine (*Materials and Methods*). It is noted that most of these samples are in nano-dimensional forms, either as nanoparticle (0-dimension), nanowire (1-dimension) or nanosheet (2-dimension). Figure 2(a) show a typical SAED pattern of a sample from the K extraction of a $K_2Fe_4Se_5$ crystal, with superstructure spots $\sqrt{5} \times \sqrt{5} \times 1$. The frequently observed twinned superstructure in $A_2Fe_4Se_5$ was also found in these samples and nanosheets from hydrothermal process, as shown in Fig. 2(b) and (c). The difference between the SAED patterns in Fig. 2(b) and (c) is the systematic absence in ($h00$), ($0k0$), $h$ odd, $k$ odd, shown by a blue arrow. The appearance of this systematic absence has been attributed to three possibilities: (*i*) potassium vacancy orders (12, 13), (*ii*) phase transformation due to sample oxidation (29), and (*iii*) Fe-Fe paring (dimer formation) (5). The former two are less possible in our case since no potassium is in the sample, and the last one was known to appear at a very low temperature, i.e., 11 K. An alternative explanation for the appearance of these forbidden spots will be given when Fe-vacancy order in $\beta$-Fe$_3$Se$_4$ is introduced.

Figure 2(e) shows the temperature dependence of magnetic susceptibility of a collection of $\beta$-Fe$_4$Se$_5$ nanosheets. The zero-field-cooled (ZFC) curve reveals a broad transition around 100 K, consistent with the resistivity data from a $\beta$-Fe$_4$Se$_5$ sheet (Fig. 2f), where at low temperatures (2–45 K), the data can be best fit with a metal-insulator transition following the 3D Mott variable range hopping model. It is interesting to note that the susceptibility of the $\beta$-Fe$_4$Se$_5$ sheets from the K$_2$Fe$_4$Se$_5$ crystal show strong orientation dependence, as shown in Fig. 2(g). Our observations clearly indicated that $\beta$-Fe$_4$Se$_5$ is an antiferromagnetic insulator at low temperatures, similar to that of the K$_2$Fe$_4$Se$_5$ (12, 17). The estimated magnetic moment from the susceptibility data (left inset of Fig. 2e) above 200 K is about 0.003 $\mu_B$ using the Curie-Weiss fit. A detailed magnetic structure of the $\beta$-Fe$_4$Se$_5$ sample is currently under investigation using neutron scattering. We further found that by annealing the $\beta$-Fe$_4$Se$_5$ nanosheets at 700 ºC in vacuum and quenching to room temperature, it became superconducting with a $T_c$ onset of 8.5 K (right inset of Fig. 2e), similar to that of the $\beta$-Fe$_{1+\delta}$Se.

More surprises arose when we looked into details of additional synthesized samples. Besides the $\sqrt{5} \times \sqrt{5} \times 1$ Fe-vacancy order, we observed several different types of Fe-vacancy order in tetragonal $\beta$-Fe$_{1-x}$Se. The second type of Fe-vacancy order we observed is shown in Fig. 3. We recorded five zone-axis patterns (two were shown in Fig. 3 and the others in Fig. S1) with 12 measured interplanar spacings (Table S1) and four angles between zone-axis patterns (Table S2). They are very close to those from superconducting tetragonal $\beta$-Fe$_{1+\delta}$Se, yet not exactly the same. The tilting procedure during SAED patterns acquirement (shown in Fig. S2a) and the angles between zone-axis patterns summarized in Table S2 also indicate that it is closer to a tetragonal phase rather than other phases in Fe-Se phase diagram (3). We refined the lattice parameters using a tetragonal lattice yielding $a = 3.67(1)$ Å and $c = 5.70(3)$ Å, about 3% shrinkage along $a$ and $b$ axes and 3% elongation along $c$-axis, compared with the (bulk) tetragonal $β$-Fe$_{1+δ}$Se ($a = 3.775$ Å and $c = 5.527$ Å) (3). However, we could not assign the presence of extra rows of spot in the [-121] zone-axis pattern (Fig. 3b, indicated by white arrows) to any of the indexes or double diffractions in tetragonal $β$-Fe$_{1+δ}$Se. In fact, these superlattice spots are too strong, making them less likely to be the result of either symmetry breaking by atomic displacements in FeSe (5) or charge density wave (CDW) (30) at low temperatures. We thus considered the modulation of atoms such as interstitials and vacancies of Fe and Se. Detailed structure calculations and simulations allowed us to exclude the ordering of Fe interstitials and Se vacancies (see Supplementary Information for details). Our calculation suggested a superstructure of 2 × 2 with a $d_{100}$ shift every other (001) plane (the FeSe$_4$ tetrahedron layer), as illustrated in Fig. 3(c). This superstructure generated diffraction patterns that are in excellent agreement with those experimental SAED patterns. This phase was denominated as $β$-Fe$_3$Se$_4$ ($x = 0.25$), and we found this phase in two different samples prepared by a high pressure route (Fig. 3) and a hydrothermal chemical route (Fig. S3). However, it is noted that the $β$-Fe$_3$Se$_4$ phase in Fig. 3 is not an unique solution for these SAED patterns. For example, the superstructure in Fig. 5(e) which shows a $\sqrt{2} \times 2\sqrt{2}$ order could as well generate the same SAED patterns, with doubled modulation along <111> direction. Because the particle investigated in this study was along [-131] direction when the sample holder was inserted, it was not possible to tilt to the $c$ axis where the aforementioned two types of order could be easily distinguished. Another feature to note is that the $c$ axis zone pattern of $β$-Fe$_3$Se$_4$ is similar to that of the superconducting $β$-Fe$_{1+δ}$Se, except that the systematic absences in ($h$00), (0$k$0), $h$ odd, $k$ odd, are now visible (Fig. S4). It is thus possible, that the SAED patterns obtained in Fig. 1(b) and (c) are without and with $β$-Fe$_3$Se$_4$ as a second phase in the field of study, respectively. It was reported that the 2 × 2 (× 1) square Fe-vacancy order (Fig. 5c) was shown theoretically to not be energetically-favorable compared with the rhombus order (2 × 4, Fig. 5d) (31). In fact, without considering the magnetic moments of Fe (nonmagnetic state), the square ordered structure has lower energy than the rhombus ordered pattern (31), though the rhombus order would become a more favorable state with an A-collinear AFM order (31). A more detailed theoretical study will clarify if the in-plane square order with a $d_{100}$ shift every other plane with AFM orders would become more favorable.

The third type of Fe-vacancy order is illustrated in Fig. 4 (crystal structure parameters of the three observed phases are summarized in Table S4). Five zone-axis patterns (two were shown in Fig. 4 and the others in Fig. S5) were recorded, and the fundamental Bragg spots yielded a tetragonal lattice with $a$ = 3.76(1) Å and $c$ = 5.47(3) Å, about 1% shrinkage along $c$ axis, as compared to the tetragonal $\beta$-Fe$_{1+\delta}$Se ($a$ = 3.775 Å and $c$ = 5.527 Å) (3). Although we could not record the $c$ axis zone pattern due to the limitation of the double tilt holder, we can still resolve the superstructure by the unique modulation wave vectors $\mathbf{q}_4$ = (2/5, 1/5, 0) and $\mathbf{q}_5$ = (1/5, 2/5, 0), as shown in Fig. 4(a) and (b), respectively. A twinned Fe-vacancy order of $\sqrt{10} \times \sqrt{10}$ with a ½$d_{310}$ shift every other (001) plane yielding an I-cell for vacancies is realized as illustrated in Fig. 4(c) and is thus assigned as $\beta$-Fe$_9$Se$_{10}$ ($x$ = 0.1). This superstructure has not been observed in A$_{1-x}$Fe$_{2-y}$Se$_2$, although it is not necessary to be unique in $\beta$-Fe$_{1-x}$Se.

Figure 5 shows a number of Fe-vacancy orders possibly observed in $\beta$-Fe$_{1-x}$Se, from $x$ = 0.5 to $x$ = 0.1, with the vacancies distributed as uniformly as possible in the iron plane. We believe that under suitable conditions, other vacancy orders could possibly exist as well. It is surprising to see that, through unconventional routes, a PbO-type tetragonal $\beta$-Fe$_{1-x}$Se compound exists in a composition range far beyond the previous reported range (2-4) to a region with a great deficiency of iron (maximum $x$ = 0.25 thus far). Thus, we need to modify the Fe-Se binary phase diagram. Moreover, our finding not only reflects the fact that the previous arguments on the superconductivity in the Fe-

Se system may be incomplete but also provides an alternative route to dope or to tune the Fe-based superconductors. In fact, when examining in more detail previous literature, we found that in MBE-grown ultra-thin FeSe films the superconductivity could be completely destroyed by Se doping (32). The films were semiconducting where the extra Se dopants were ordered into a $\sqrt{5} \times \sqrt{5}$ superstructure at a high doping concentration (figure S5 in ref. (32)). Such an observation is consistent with the presence of Fe vacancies in a Se-rich growth environment and ordering into superstructures in these ultra-thin FeSe films.

Based on the observation in this study, we proposed a temperature-doping phase diagram for Fe-Se superconducting system, as illustrated in Fig. 6, which is very similar to the phase diagram of the superconducting $La_2CuO_{4+y}$ (33). The magnetic and insulating/semiconducting phases of $\beta$-$Fe_{1-x}$Se with Fe-vacancy orders may act as the parent phase of FeSe superconductor, instead of the previously argued $\beta$-$Fe_{1+\delta}$Te (6), which shows different magnetic and electronic features compared with $\beta$-$Fe_{1+\delta}$Se (25, 34-36). The superconducting $\beta$-$Fe_{1.01}$Se with $T_c$ = 8.5 K could in fact be in the overdoped regime in our proposed phase diagram, because we showed that it could result from the $\beta$-$Fe_4Se_5$ nanosheets by annealing at 700 °C in vacuum and then quenching to room temperature. Further introduction of Fe suppresses both the tetragonal-to-orthorhombic structural transition and superconductivity as suggested by Mcqueen et al. (5) $\beta$-$Fe_{1-x}$Se with vacancy orders could exhibit particular AFM orders at a low temperature and are not superconducting. Doping or suppression of AFM orders may initiate superconductivity. In fact, recent experimental (37, 38) and theoretical (39) evidence suggested that disorder of Fe vacancies or suppression of the Fe-vacancy orders may bring the reemergence of superconductivity in $K_{1-x}Fe_{2-y}Se_2$. Efforts to examine the correlation between the suppression of Fe-vacancy orders and superconductivity in $\beta$-$Fe_{1-x}$Se are currently in progress.

In summary, we report the discovery of the presence and ordering of iron vacancies in PbO-type tetragonal $\beta$-Fe$_{1-x}$Se. Analytical TEM indicates that at least three types of Fe-vacancy order were found: $2 \times 2$ with a $d_{100}$ shift every other (001) plane, namely $\beta$-Fe$_3$Se$_4$ ($x = 0.25$), $\sqrt{5} \times \sqrt{5} \times 1$, namely $\beta$-Fe$_4$Se$_5$ ($x = 0.2$), and $\sqrt{10} \times \sqrt{10}$ with a $\frac{1}{2}d_{310}$ shift every other (001) plane, namely $\beta$-Fe$_9$Se$_{10}$ ($x = 0.1$), similar to the recently discovered Fe-vacancy orders in A$_{1-x}$Fe$_{2-y}$Se$_2$ (A = K, Rb, Cs, Tl). No superconductivity was observed in $\beta$-Fe$_4$Se$_5$. Instead, a broad magnetic transition was found around 100 K. This discovery suggests that the rich-phases found in A$_{1-x}$Fe$_{2-y}$Se$_2$ may not be exclusive cases in iron-based superconductors. Furthermore, the magnetic and semiconducting/insulating $\beta$-Fe$_4$Se$_5$ or other $\beta$-Fe$_{1-x}$Se phases with a particular Fe-vacancy order may serve as the parent phase of the Fe-Se superconducting system instead of the previously assigned $\beta$-Fe$_{1+\delta}$Te. Our finding creates interests for future test, both experimentally and theoretically, on the role of Fe vacancies or dopants in iron-based superconductors.

**Materials and Methods**

**Crystals**. FeSe crystals were synthesized via a high pressure route: Fe and Se powders (3N purity) with appropriate ratio (1.03:1 in atomic percentage) were mixed and cold-pressed into pellets. The pellet was placed in a BN crucible mounted inside a pyrophyllite cube with a graphite heater, assembled in a wedge-type cubic-anvil high-pressure apparatus (model TRYCA 350, SANKYO Co. Ltd., Osaka, Japan). A pressure of ~2.8 GPa was applied at room temperature and throughout the whole process. The temperature was ramped up to 1050 ˚C and kept for 1.25 hours to form the melt. Then it was cooled within 30 minutes to 850 ˚C and quenched to 350 ˚C for in situ annealing. After 30 minutes annealing, it was quenched to room temperature.

**Crystal-like Sheets**. For small $c$ axis-oriented $\beta$-Fe$_4$Se$_5$ sheets, K$_x$Fe$_{1.6+y}$Se$_2$ crystals were firstly made using Bridgman method with a starting composition of K$_{0.8}$Fe$_2$Se$_2$ (40). Potassium in these $c$- axis-oriented pieces of $K_xFe_{1.6+y}Se_2$ crystal was removed by $I_2$ in acetone. Pieces of the resulted β-$Fe_4Se_5$ crystal-like sheets were finally obtained by repeated cleaving using Scotch-tape method.

**Nanosheets**. Se nanoparticles were synthesized prior to the formation of FeSe nanosheets. 80 ml of ethylene glycol was added into a three-neck flask equipped with a standard Schlenk line, together with PVP, NaOH and $SeO_2$ powder with a ratio of 1 : 3 : 1.66 (in molar ratio). The mixture was then heated to 160 ºC, followed by adding 2.4 ml of hydrazine hydrate as reducing agent. The Fe precursor solution was made by dissolving 2.3857 g of $FeCl_2$ in 20 ml ethylene glycol. It was then injected into the above mixture at 160 ºC and kept for 12 hours for the formation of FeSe nanosheets. To dry the products, we first remove the capping ligands on FeSe nanosheets by dispersing in acetone with diluted ethanol (10% vol/vol) and stirring until they were precipitated. The supernatant was decanted and the precipitates were washed with acetone for three times to remove the residual PVP. The nanosheets were finally collected by centrifugation and dried in vacuum.

**Nanowires**. FeSe nanowires were grown via a two-step process, called the on-film formation of nanowires (OFF-ON) method (41). FeSe thin films were firstly deposited (42) and then annealed at 400 ºC for 120 hours in vacuum.

**Structure Characterization**. TEM samples of FeSe crystals and sheets were made by smashing a small piece of cleaved crystal immersed in ethanol in a mortar, and then collected using Lacey carbon grids. FeSe nanosheets were suspended in ethanol and collected by Lacey carbon grids. Samples of FeSe crystals were analyzed by a JEOL 3000F microscope equipped with an INCA EDX system. Samples of FeSe nanosheets and nanowires were analyzed by a JEOL 2100F microscope equipped with an OXFORD X-Max EDX analyzer. Lattice constants were calculated from interplanar spacings in the SAED patterns and refined by a non-linear least squares cell refinement program (43). Simulations of electron diffractions were carried by MacTempas (http://www.totalresolution.com/MacTempasX.htm).

**Physical property Measurements**. The resistances of small pieces of FeSe sheets and FeSe nanowires were measured with a PPMS system (Quantum Design Inc.) using 4 point and 2 point method, respectively. The magnetic properties were measured on a SQUID magnetometer (MPMS, Quantum Design Inc.).

Acknowledgements

We thank Dr. Phillip Wu for his valuable suggestions and Dr. T. K. Lee for fruitful discussions. Technical support from NanoCore, the Core Facilities for Nanoscience and Nanotechnology at Academia Sinica in Taiwan, is acknowledged. This work was supported by the National Science Council, the Academia Sinica of Taiwan, and the US AFOSR/AOARD.

Author Contributions

T. K. Chen, C. C. Chang, and C. M. Tseng performed the TEM experiments. T. K. Chen and H. Y. Tang performed the analysis and simulation of electron diffractions and built the models. C. C. Chang, A. H. Fang, C. H. Wang, W. H. Chao, and Y. R. Wu prepared the samples. H. H. Chang, A. H. Fang, Y. C. Lee, M. H. Wen performed the transport measurements. A. H. Fang and C. H. Wang performed the magnetic property measurements. F. R. Chen and D. Van Dyck assisted with the TEM experiments. M. J. Wang assisted with the transport measurements. T. K. Chen and M. K. Wu wrote the paper. M. K. Wu conceived the idea and supervised the work.

Competing Financial Interests

There is no competing financial interest.

**Figure Legends**

Figure 1. (a) TEM image of a FeSe nanowire. (*Inset*) Temperature-dependent transport property of the same nanowire. (b) The SAED pattern of the nanowire, revealing a tetragonal lattice along the [001] zone-axis direction. Superstructure wave vectors $\mathbf{q}_1 = (1/5, 3/5, 0)$ and $\mathbf{q}_2 = (3/5, 1/5, 0)$ are indicated by red arrows.

Figure 2. Experimental SAED patterns of (a) untwinned and (b and c) twinned $\beta$-Fe$_4$Se$_5$. Superstructure wave vectors $\mathbf{q}_1 = (1/5, 3/5, 0)$ and $\mathbf{q}_2 = (3/5, 1/5, 0)$ are indicated by red arrows. The blue arrow in (c) points out one of the systematic absence in ($h$00), (0$k$0), $h$ odd, $k$ odd, which is not visible in (b). (d) Schematic drawing of atom arrangements for untwinned and twinned $\beta$-Fe$_4$Se$_5$. (e) Temperature-dependent susceptibility of a collection of $\beta$-Fe$_4$Se$_5$ nanosheets. (*Left inset*) Curie-Weiss fit in the high temperature range 200 K < $T$ < 300 K: $\chi = C/(T-\theta) + \chi'$, where $C$ is the Curie constant, $\theta$ is the Curie temperature, and $\chi'$ is the temperature-independent susceptibility. (*Right inset*) Temperature-dependent susceptibility of the same $\beta$-Fe$_4$Se$_5$ nanosheets after annealed at 700 ºC in vacuum and quenched to room temperature. (f) Resistance as a function of temperature for three $\beta$-Fe$_4$Se$_5$ sheets, cleaved by scotch-tape from different pieces in the same batch. Current was applied within the *a-b* plane. (*Inset*) Natural logarithmic resistance plotted against $1/T^{1/4}$, in which the data between 2 and 45 K are fit by the 3D Mott variable range hopping model: $R \sim \exp\left[-(T^*/T)^{1/4}\right]$, where $T^*$ is the characteristic temperature. (g) Temperature-dependent susceptibility of a $\beta$-Fe$_4$Se$_5$ sheet. Magnetic field was applied along and perpendicular to the sheet normal, i.e., the *c* axis of the tetragonal lattice.

Figure 3. From *Left* to *Right*: experimental SAED and simulated kinematical electron diffraction (KED) patterns of β-Fe$_3$Se$_4$ and β-FeSe in the zone axes of (a) [-121] and (b) [-111]. Other zone-axis patterns are shown in Fig. S1. White arrows in (a) indicate extra rows of spots compared with the KED pattern of β-FeSe. Superstructure wave vector **q**$_3$ = (1/2, 1/2, 1/2) is indicated by a red arrow. Crosses in the KED patterns of β-FeSe denote double diffraction spots. (c) Schematic drawings of atom arrangements for β-Fe$_3$Se$_4$ from different viewing directions. Solid lines enclose the I-cell for Fe vacancies.

Figure 4. (*Left*) Experimental SAED and (*middle*) simulated KED patterns of twinned β-Fe$_9$Se$_{10}$ in the zone axes of (a) [-212] and (b) [-121]. Other zone-axis patterns are shown in Fig. S5. Superstructure wave vectors **q**$_4$ = (1/5, 2/5, 0) and **q**$_5$ = (2/5, 1/5, 0) are indicated by blue arrows. The twinned supersturctural reflections are illustrated by blue and red circles. (c) Schematic drawings of atom arrangements for twinned and untwinned β-Fe$_9$Se$_{10}$. Solid lines enclose the I-cell for Fe vacancies.

Figure 5. Possible Fe-vacancy orders in β-Fe$_{1-x}$Se and their corresponding stoichiometry. (a) $x$ = 0.5; (b) $x$ = 0.333; (c–e) $x$ = 0.25; (f) $x$ = 0.2; (g and h) $x$ = 0.167; (i) $x$ = 0.143; (j and k) $x$ = 0.125; and (l) $x$ = 0.1. Se atoms were hidden for better visibility. Solid and open circles denote Fe atoms and Fe vacancies, respectively. Green area encloses the original β-FeSe unit cell. Black lines connect nearest Fe vacancies to improve the visibility of the order pattern.

Figure 6. Sketch of the proposed temperature-doping phase diagram of Fe-Se superconducting system, showing regions for antiferromagnetism (AFM) and superconductivity (SC).

(a) 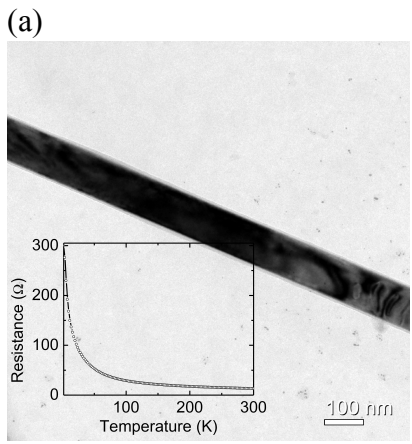 (b) 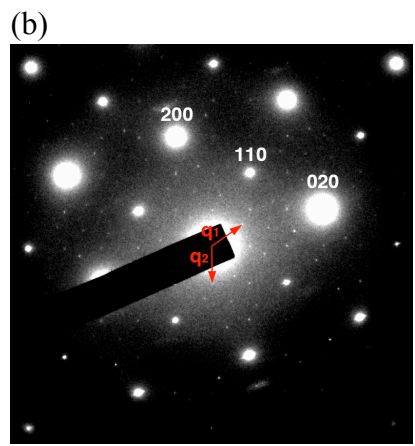

Figure 1

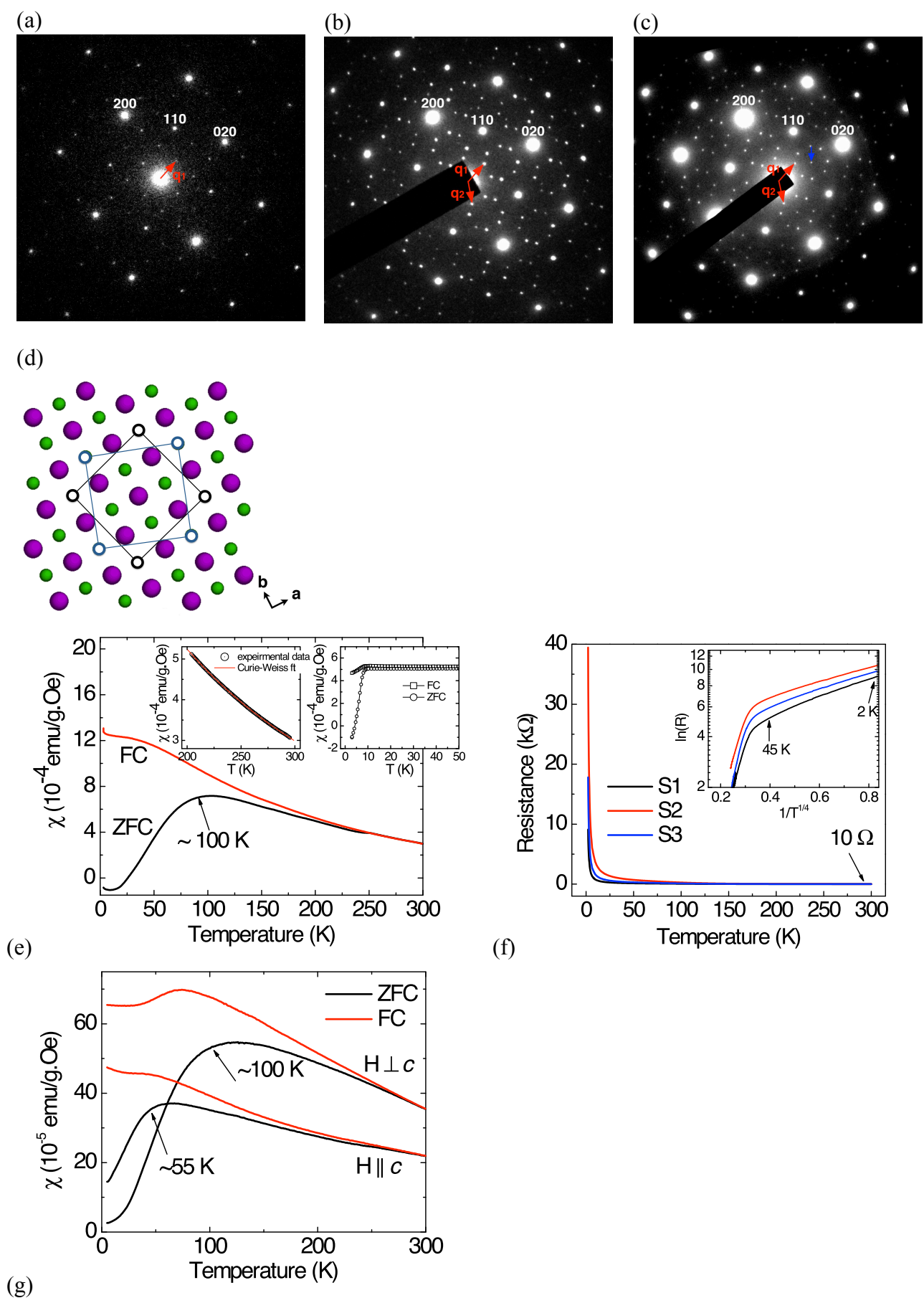

Figure 2

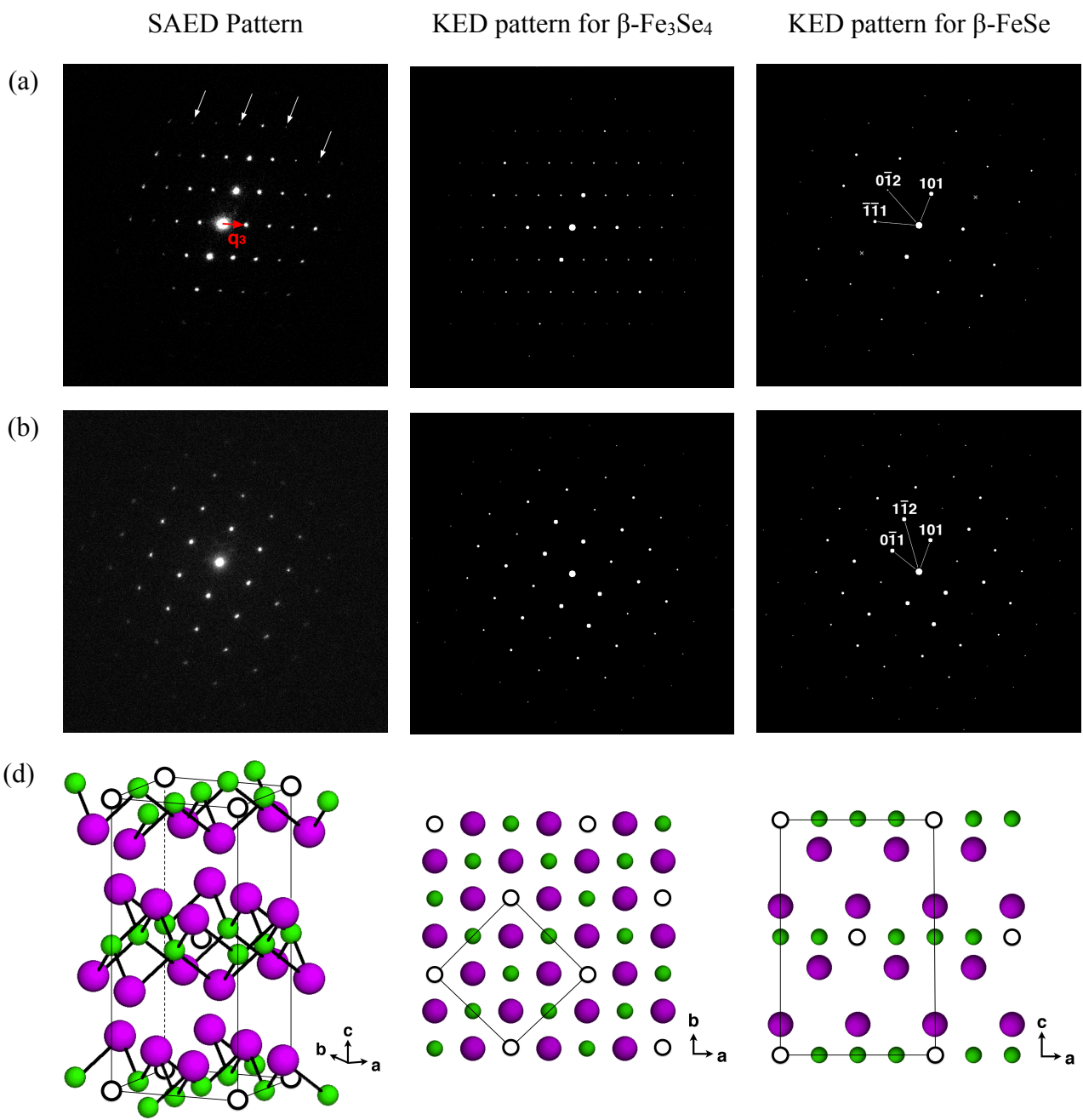

Figure 3

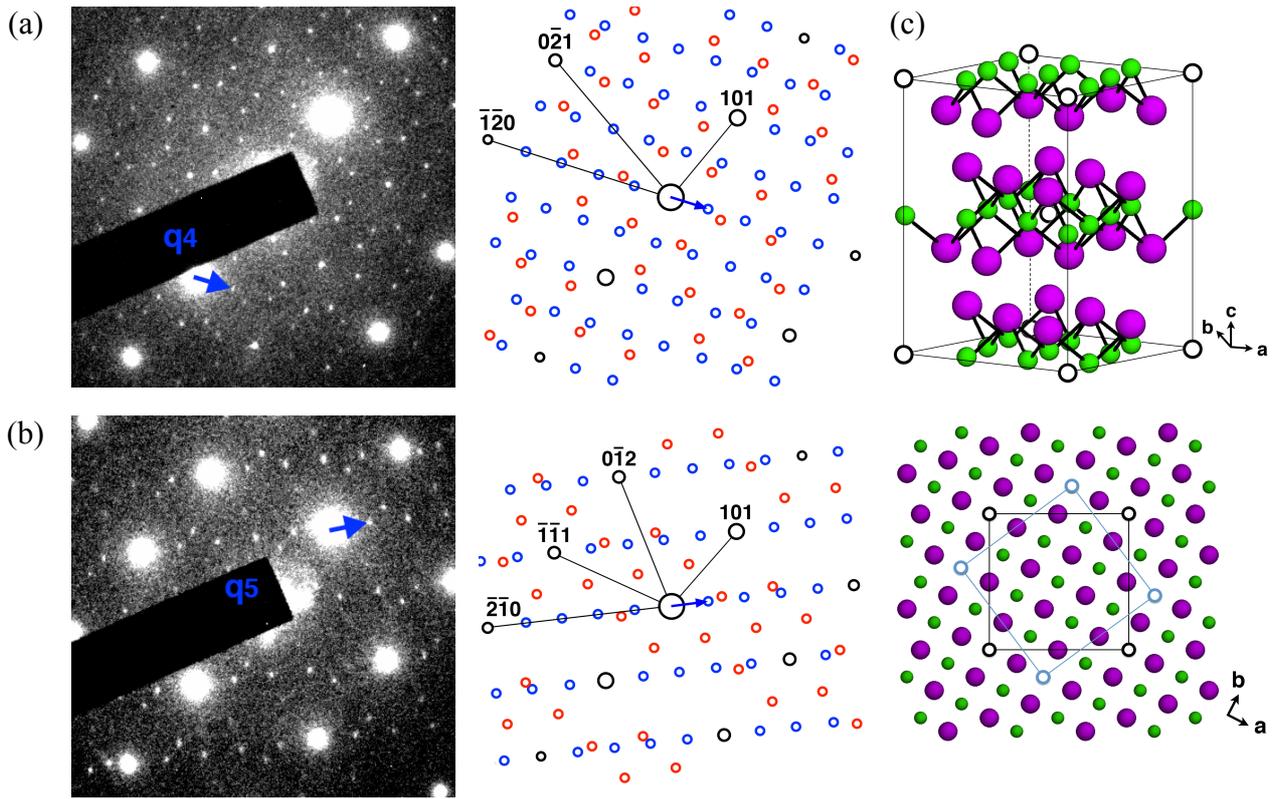

Figure 4

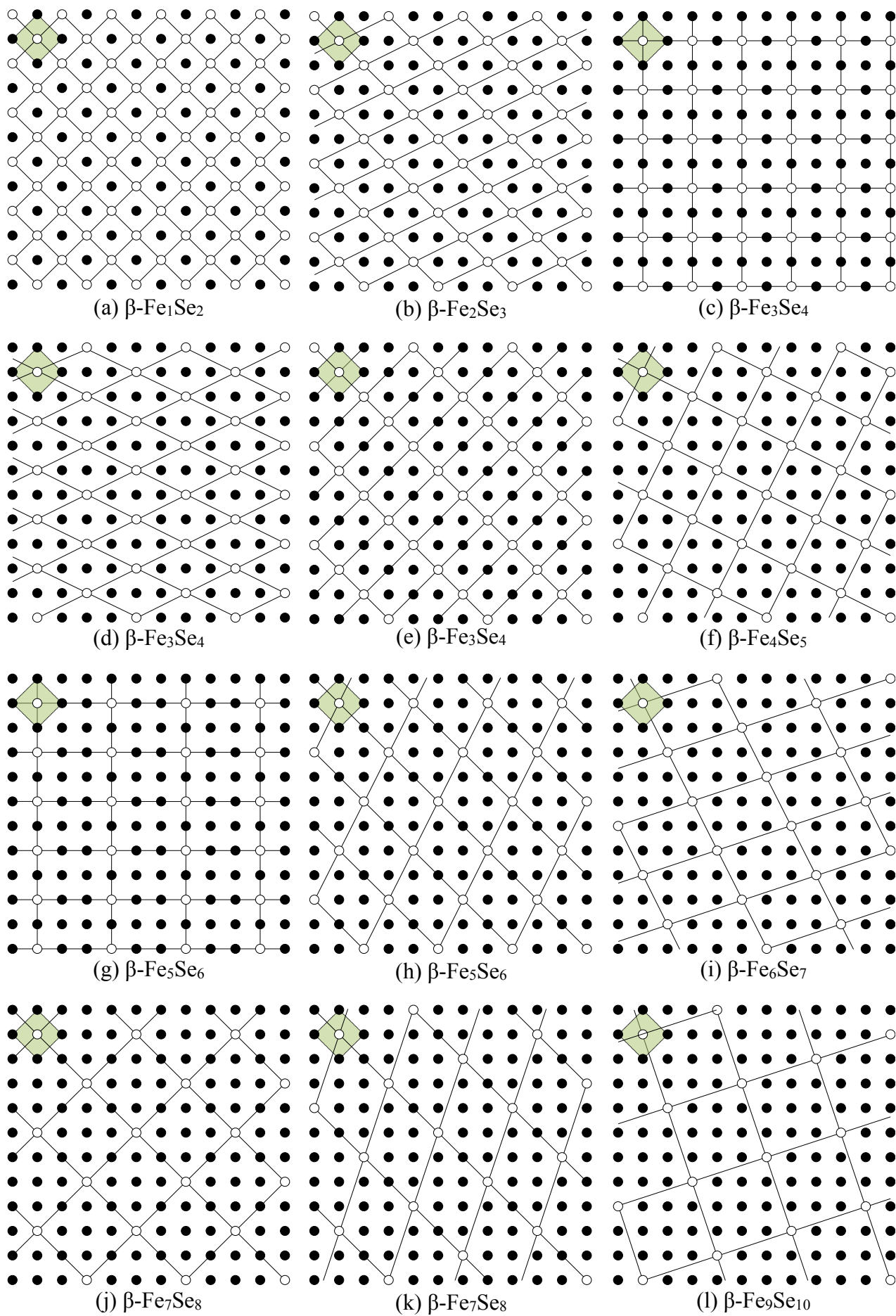

Figure 5

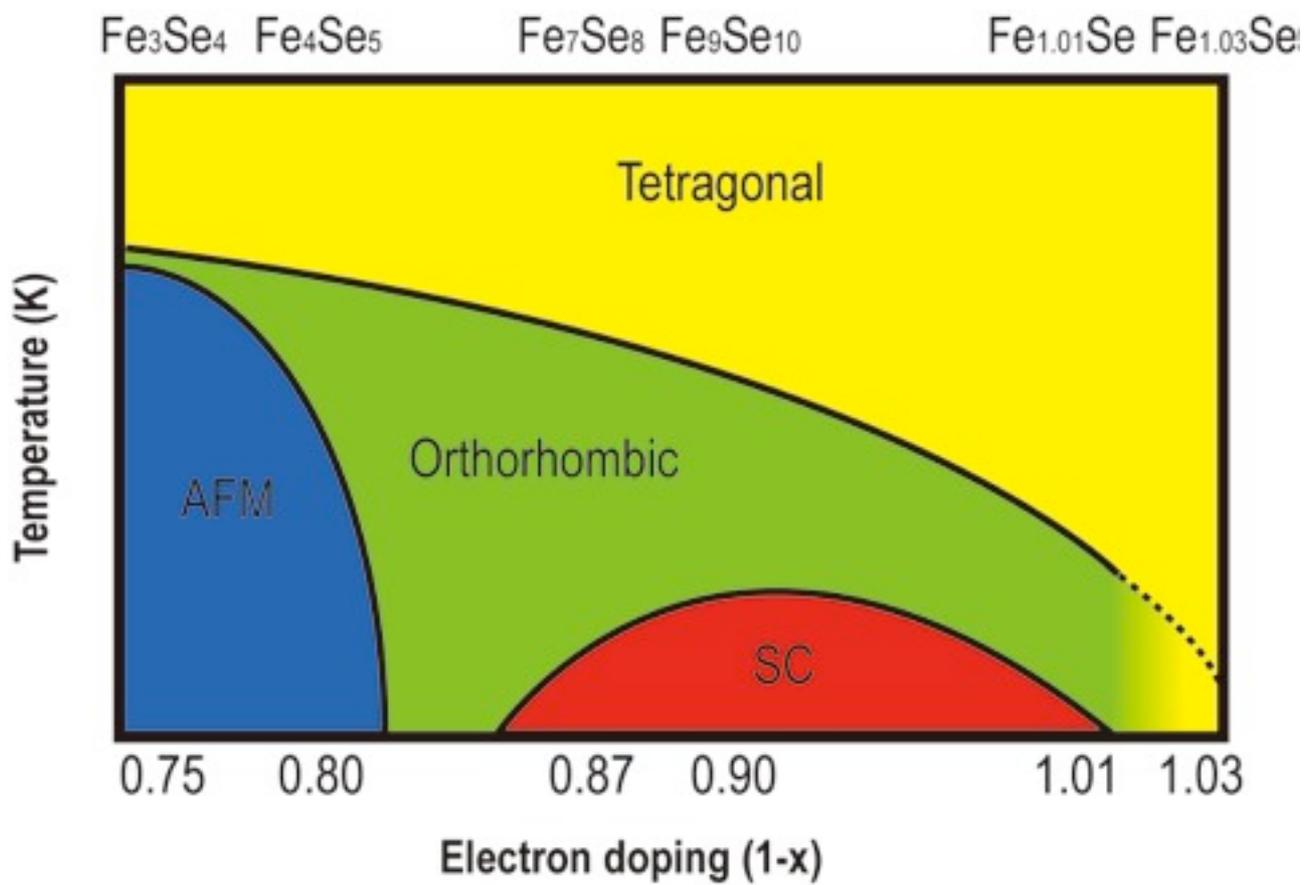

Figure 6